# An Evaluation of Arabic Language Learning Websites

Hadhémi ACHOUR
Tunis High School of Management (ISG Tunis)
University of Tunis
Tunisia
Hadhemi_Achour@yahoo.fr

Wahiba Ben Abdesslam
Tunis High School of Management (ISG Tunis)
University of Tunis
Tunisia
wahiba.abdessalem@isg.rnu.tn

*Abstract—* As a result of ICT development and the increasingly growing use of the Internet in particular, practices of language teaching and learning are about to evolve significantly.
Our study focuses on the Arabic language, and aims to explore and evaluate Arabic language learning websites. To reach these goals, we propose in a first step, to define an evaluation model, based on a set of criteria for assessing the quality of websites dedicated to teaching and learning Arabic. We subsequently apply our model on a set of Arabic sites available on the web and give an assessment of these web sites. We finally discuss their strengths and limitations. (Abstract)

*Keywords—* Web-based language learning; language learning websites; Arabic language learning; learning website evaluation;

## I. Introduction

Internet is today a huge international network that is being used in all areas of life. In the field of education in particular, the contributions of the Internet are very important especially in regards to distance education and autonomous learning [Kartal 2010]. A wide variety of educational web applications are indeed available, allowing learners to overcome the constraints of time, distance and boundaries. We cite for example: e-learning platforms (LMS: Learning Management System), learning portals, learning object warehouses, blogs, wikis, virtual communities and educational social networks, ...

Web-based language learning is language learning that involves the use of the Web and exploits Web materials, resources, applications or tools [Son 2007]. In this paper, we focus on Arabic language learning websites. Our aim in this study, is to explore and evaluate websites dedicated to teaching and learning Arabic. To reach these goals, we propose to define an evaluation model, based on a set of criteria for assessing the quality of such websites. We subsequently apply our model on a set of Arabic sites available on the web and give an assessment of these web sites. We finally discuss their strengths and limitations.

## II. Language Learning Websites

Several authors agree that the Internet has major potential in teaching and learning languages [Kartal 2005] [Wang 2009]. The Web in particular, offers a global database of authentic materials that can enhance language learning and teaching [Son 2005]. A large number of various language learning websites are nowadays available on the web. [Kartal 2005] has proposed a simple classification that distinguishes between two types of language learning sites: those designed for the purpose of language skills (reading, writing and listening) and those concerned with language domains (grammar, phonetics, vocabulary, Culture and civilization).

We should however add that, not all materials are equally reliable or valuable [Son 2005], and language learning websites are not always of good quality. On this, [Kartal 2005] says that most language learning websites don't include all opportunities that the Internet provides. He adds that almost all of these learning sites offer a limited pedagogical approach which is always reduced to just answering structural auto-corrective exercises (such as multiple choice questions, true or false items, and fill in the blanks). Still according to Kartal, in these sites, pedagogical scenarios and learning theories are not reflected and objectives, levels and the target audience are not indicated [Kartal 2005].

Hence, language learning website assessment, based on appropriate quality criteria, becomes necessary to guide developers in designing and creating these sites, and to guide both teachers and learners in their quest for useful and reliable sites that meet their needs.

## III. Learning Website Evaluation: A Literature Review

Several studies have focused on evaluating language learning websites. For that, they have defined a set of criteria to assess the quality of these sites. [Nelson 1998] suggests a system for evaluating ESL websites (English as a second language websites), which is divided into four parts: *Purpose* (intended goals, uses, and audiences), *Pedagogy* (instructions, aspects of multimedia, interactivity, communicativeness), *Design/Construction* (general web design principles such as appearance, navigation, load speed, ...), and *Description/Other* (general description and relevant comments about the site).

[Kelly 2000] proposes the following list of points that should be considered when designing a website for ESL (English as Second Language)

students: a) *usability* by a wide audience as possible, b) *speed* of loading and displaying, c) *ease of use* (ease of navigation and reading), d) *usefulness* (the site should fill a need), e) *integrity* and *professionalism* (honesty, accuracy, respect of copyrights, indicating the date of last update, a contact address, …), f) *wise and effective use of "cutting edge technology"*, g). [Kelly 2000] also recommends to make the site friendly and fun to use, and to worry about the minority who use less powerful computers, older browsers and have slow Internet access.

[Son 2005] presents a model for Web site categorization and evaluation, and reports the results of a review of selected English as a second/foreign language (ESL/EFL) Web sites using this model. Son's model is based on 15 criteria covering: a) *Purpose* (is the purpose clear? is the content in line with the purpose?), b) *Accuracy* (is the content accurate?), c) *Currency* (is the Website current? is the Website updated regularly?), d) *Authority* (is there information on the author?), e) *Loading speed*, f) *Usefulness* (does the Website provide useful information and language activities?), g) *Organization* (is the Website well organized and presented?), h) *Navigation* (is the Website easy to navigate?), i) *Reliability* (is the Website free of bugs, dead links, breaks?), j) *Authenticity* (are the learning materials authentic?), k) *Interactivity*, l) *Feedback*, m) *Multimedia*, n) *Communication* (can the user communicate with real people on-line through the Website?), o) *Integration* (can the learning materials be integrated into a curriculum? does the content fit with curricular goals?). To these criteria, a site reviewer highlights the site to be "Very Unsatisfactory", "Unsatisfactory", "Uncertain", "Satisfactory" or "Very Satisfactory" [Son 2005].

The work presented by [Kartal 2010] adresses constructing foreign language learning websites, and proposes a set of characteristics of a good website:

- *physical characteristics* that are mainly related to the website design and constitute its general structure. They include an appropriate choice of colors, a clearly organized parts and sections with easy transition between them, the ability to use online dictionaries or some other programs in concordance with the site activities and exercises, the ability to find various materials related to linguistic subjects or skills;

- *Contextual Characteristics* that mainly relate to the features of the site content. The author [Kartal 2010] says indeed, that available materials should be of every type (written, visual, audio), appropriate for the concerned level, subject, or type. They should be up to date and authentic and supported by extrinsic programs and tools (such as search engines, newspapers, magazines, …). Learners should be able to access customized resources and get feedback on their activities.

- *Pedagogical Characteristics* that contribute to the learning and teaching process regarding the use of educational methods and approaches.

For the purpose of analyzing and evaluating a grammar website, [Sabri 2009] proposes a mixed evaluation approach, that relies on both, a grid of evaluation criteria and a practical usability test of the website, referring to 2 procedures: heuristic and empirical. The main components considered in this evaluation approach are: *website description*, *website ergonomy* (interface, navigation; learning path), *usability test* (type of difficulties encountered while running tasks), *complementary tools* (Dictionaries, translators, …).

We should also note that several assessment grids of language learning sites were developed and are available online. Among these grids, we can cite the example of Perrot[1], who proposes to take into account three aspects: *site presentation* (general features of the site), *interface analysis* (quality of content and quality of navigation) and *pedagogical analysis* (analysis of activities, types of exercises and error processing models).

The models here above introduced, are different, but converge on many aspects and have in fact, much in common: they all take into consideration - each in its way and with different degrees of detail- the main principles of ergonomics and HCI (Human Computer Interaction), educational aspects and elements of interactivity.

These models may complement each other, and may also be enriched by additional quality criteria. In the following section, we propose an evaluation model that we plan to use in order to review a set of Arabic language websites. This would allow us to contribute to a state of the art development, regarding the Arabic language learning websites presently available on the web.

IV. AN EVALUATION MODEL FOR ASSESSING ARABIC LANGUAGE LEARNING WEBSITES

To build our assessment model, we relied on both an exploratory research of current language learning websites, and on the literature review of works relating to the assessment of such kind of sites.

The model we propose in this work, is a general model, in which we tried to take into account various general principles related to ergonomics, linguistic and pedagogical aspects. At present, we do not distinguish between websites which are specifically dedicated to foreign language learning (FL), second foreign language learning (SFL) or native speaker language (NSL) learning, even though pedagogical approaches differ from situation to another. The proposed model is general and therefore, can be used for assessing different types of language learning websites. We plan in a future work, to specify the model and adapt it to different learning situations depending on whether the target language is either FL or SFL or NSL.

---

[1] http://www3.unileon.es/dp/dfm/flenet/grilles2.html#T Perrot. Retrieved on May 28 2012.

The proposed model draws on works presented in the previous section (literature review) and particularly, on the works of [Son 2005] and [Kartal 2010]. In this model we suggest to:

- Reorganize differently, the evaluation criteria into a more general model that could be used to evaluate any kind of language learning websites.
- Enhance previous models by additional criteria. Indeed, we expand criteria related to the communication aspect, since the various communication forms (chat, forum, discussion group, social networks) can support language learning through social interaction. We also take into account more criteria referring to learning personalization, and emphasize more on the utility of natural language processing tools and linguistic resources. In addition, we take into account the users opinions and ratings of the websites according to they are students or teachers.

Our model is thus organized into nine sections as follows:

*A. General website Information*

This section lists a set of relevant and helpful information that should be clearly specified on the learning site such as : **target audience** (general/ specific group), **target level** (elementary / intermediate / advanced), **target language learning** (FL / SFL / NSL), **interface language** (monolingual / multilingual), **date of website creation**, **date of last update**, **author identity**, **contact address**, **information on users** (number of visitors / number of registered users, …), **site rating by learners**, **site rating by teachers**, **users comments** , ...

*B. Language Skills and Fields*

Within this model component, we specify the purpose of the website in terms of language skills, activities and areas of language: ***Reading***, ***Writing***, ***Listening***, ***Speaking***, ***Grammar***, ***Phonetics***, ***Vocabulary***, ***Other***.

*C. Educational material*

This section is about the educational materials that are available on the website: ***lessons***, ***exercises***, ***educational games***, ***testing tools***, …).

*D. Multimedia Use*

This section includes the different types of educational materials: ***text***, ***graphics***, ***sound***, ***video***.

*E. Interactivity*

Interactivity component aims to assess the learning site interactivity by identifying ***activities feedback***, ***pedagogical guidance and explanations***, ***customized guidance and explanations***, …

*F. Communication.*

This section is about the different communications forms offered by the learning site: ***chat***, ***discussion forums***, ***e-mail***, ***social network***, …

*G. Aid tools and linguistic resources*

This component covers natural language tools and software, and linguistic resources provided by the website and which can be very helpful for learners in their learning process: ***Monolingual Dictionaries***, ***Multilingual Dictionaries***, ***Natural language processing tools*** (translators, conjugators, morphologic analysers, syntactic parsers, …), ***Search engines***, ***e-books***, ***links to other websites***, …

*H. Website Ergonomics*

This section evaluates the quality of the website ergonomics. It includes several criteria such as: ***color harmony***, ***font and legibility***, ***general structure and organization***, ***ease of navigation***, ***loading speed***, ...

*I. Content Quality*

This last section is concerned with evaluating the quality of the website content. It is composed by the following criteria: ***usefulness*** of information and language activities, content ***accuracy***, ***Adequacy*** to level and type of audience, ***reliability*** (bugs, dead links, …).

The resulting model is given in appendix 1, as an evaluation form to be completed for every reviewed language learning website. Concerning the criteria defined in sections *A* to *G*, we just check the (Yes) or (No) box, equivalent to the presence or the absence of the corresponding criterion. However, for the sections *H* and *I*, we assign to each criterion, a score from 1 to 5, corresponding to : "very poor", "poor", "medium", "good", very good".

V. EXPLORING AND EVALUATING ARABIC LANGUAGE LEARNING WEBSITES

Using the model we have defined (see appendix 1), we carried out an assessment of 10 free open access Arabic language learning websites, by filling the evaluation form for each one of these sites. Results, which are summarized in appendix 2, show that:

- regarding the general website information section of the model, the majority of sites (70%) do not specify target audience and target level. Last update dates, users information and teachers rating are also elements that are not considered by most sites.

- regarding language skills and fields, writing and listening are the major skills that are found

- in a large number of the evaluated learning websites. However, the speaking skill in Arabic language seems to be more difficult to integrate.

- in all the reviewed sites, educational materials lack exercises and evaluation tests of learners.

- multimedia (sound, video, ..) is generally well used in the majority of the evaluated sites.

- interactivity, however is a weak point in all of the reviewed Arabic learning websites. Indeed, 90% of these sites do not provide feedback on accomplished activities, and do not guide learners with explanations in their tasks.

- regarding the Communication section, the only communication form, available in almost all of the evaluated sites (90%) is the email. 80% of the sites contain neither chat nor discussion forums.

- over the 10 sites that have been evaluated, a significant number of sites (8), lack aid tools and linguistic resources (natural language processing, dictionaries, search engines, …).

- regarding the sites ergonomics, the majority of the sites quality is medium (8 per 10) and only one site has a good score for ergonomics criteria.

- the content usefulness and accuracy are of a medium quality for almost all the evaluated sites (9 part ten). The evaluation also shows that only one site is reliable with the score 'good', while 90% of these have the score 'poor' concerning their reliability.

## VI. Conclusion

In this paper, we were interested in the quality evaluation of Arabic language learning websites. For this purpose, we defined a general evaluation model, organized into 9 sections, and relying on various ergonomic, pedagogical and linguistic criteria. Using this model, we were able to make an evaluation of the current state of Arabic language learning websites. The results of this assessment has shown that the quality of these sites needs to be greatly improved, especially regarding the richness of educational materials, interactivity and reliability.

In addition, the proposed model can assist and guide both learners and teachers in their quest for suitable and reliable sites so as to improve their learning and teaching. However, and as mentioned above, the proposed model does not distinguish between websites which are specifically dedicated to foreign language learning (FL), second foreign language learning (SFL) or native speaker language (NSL) learning, even though pedagogical approaches differ from situation to another.

Therefore, we have, in a next step, to improve this model, by assigning weights to criteria depending on, whether the target language is either FL or SFL or NSL, in order to take into account that some specific criteria can be more significant (and thus have a major impact on the site quality) when the language is taught as a FL or SFL, whereas they are less important (and therefore significantly less affecting the site quality), if the language is taught as a NSL. We can cite as examples, the criteria of multilingual interface, multilingual dictionaries and translators which can be considered as being more useful and significant in the cases of SFL or FL sites.

# Appendix 1: Language Learning website Evaluation Form

| Website title | |
|---|---|
| URL | |

| 1. General website Information | | |
|---|---|---|
| | Yes | No |
| Target Audience | | |
| Target Level | | |
| Target Language Learning | | |
| Multilingual interface | | |
| Date of website creation | | |
| Date of Last Update | | |
| Author Identity | | |
| Contact Adress | | |
| Information on users | | |
| Rating by Learners | | |
| Rating by Teachers | | |

| 2. Language Skills and Fields | | |
|---|---|---|
| | Yes | No |
| Reading | | |
| Writing | | |
| Listening | | |
| Speaking | | |
| Grammar | | |
| Phonetics | | |
| Vocabulary | | |
| Other | | |

| 3. Educational material | | |
|---|---|---|
| | Yes | No |
| Lessons | | |
| Exercises: | | |
| - MCQ | | |
| - T/F Questions | | |
| - Fill in the blanks | | |
| - Drop down menus | | |
| - Click | | |
| - Drag and drop | | |
| - Other | | |
| Educational games | | |
| Evaluation test | | |
| Other | | |

| 4. Multimedia Use | | |
|---|---|---|
| | Yes | No |
| Text | | |
| Graphics | | |
| Sound | | |
| Video | | |
| Animation | | |

| 5. Interactivity | | |
|---|---|---|
| | Yes | No |
| Activities feedback | | |
| Pedagogical Guidance and explanations | | |
| Customized Guidance and explanations | | |

| 6. Communication | | |
|---|---|---|
| | Yes | No |
| Chat | | |
| Forums | | |
| e-mail | | |
| Social network | | |
| Other | | |

| 7. Aid tools and linguistic resources | | |
|---|---|---|
| | Yes | No |
| Monolingual Dictionaries | | |
| Multilingual Dictionaries | | |
| Natural language processing tools | | |
| Search engines | | |
| e-books | | |
| Links to other websites | | |

| 8. Website Ergonomics | | | | | |
|---|---|---|---|---|---|
| | 1 Very poor | 2 Poor | 3 Medium | 3 Good | 4 Very good |
| Color harmony | | | | | |
| Font and legibility | | | | | |
| General Structure and Organization | | | | | |
| Ease of Navigation | | | | | |
| Loading speed | | | | | |

| 9. Content Quality | | | | | |
|---|---|---|---|---|---|
| | 1 Very poor | 2 Poor | 3 Medium | 3 Good | 4 Very Good |
| Usefulness | | | | | |
| Accuracy | | | | | |
| Adequacy | | | | | |
| Reliability | | | | | |

# Appendix 2: Evaluation Results of 10 Arabic Language Learning Websites

Given numbers in this table are the numbers of sites that meet the specified criteria.

| 1. General website Information | | |
|---|---|---|
| | Yes | No |
| Target Audience | 3 | 7 |
| Target Level | 3 | 7 |
| Target Language Learning | 4 | 6 |
| Multilingual interface | 6 | 4 |
| Date of website creation | 5 | 5 |
| Date of Last Update | 2 | 8 |
| Author Identity | 5 | 5 |
| Contact Adress | 7 | 3 |
| Information on users | 2 | 8 |
| Rating by Learners | 4 | 6 |
| Rating by Teachers | 0 | 10 |

| 2. Language Skills and Fields | | |
|---|---|---|
| | Yes | No |
| Reading | 10 | 0 |
| Writing | 10 | 0 |
| Listening | 8 | 2 |
| Speaking | 1 | 9 |
| Grammar | 4 | 6 |
| Phonetics | 4 | 6 |
| Vocabulary | 4 | 6 |
| Other | 2 | 8 |

| 3. Educational material | | |
|---|---|---|
| | Yes | No |
| Lessons | 10 | 0 |
| Exercises: | 3 | 7 |
| - MCQ | 3 | 7 |
| - T/F Questions | 1 | 7 |
| - Fill in the blanks | 0 | 10 |
| - Drop down menus | 0 | 10 |
| - Click | 1 | 9 |
| - Drag and drop | 0 | 10 |
| - Other | 1 | 9 |
| Educational games | 2 | 8 |
| Evaluation test | 1 | 9 |
| Other | 10 | 0 |

| 4. Multimedia Use | | |
|---|---|---|
| | Yes | No |
| Text | 10 | 0 |
| Graphics | 10 | 0 |
| Sound | 10 | 0 |
| Video | 7 | 3 |
| Animation | 7 | 3 |

| 5. Interactivity | | |
|---|---|---|
| | Yes | No |
| Activities feedback | 1 | 9 |
| Pedagogical Guidance and explanations | 2 | 8 |
| Customized Guidance and explanations | 1 | 9 |

| 6. Communication | | |
|---|---|---|
| | Yes | No |
| Chat | 2 | 8 |
| Forums | 2 | 8 |
| e-mail | 9 | 1 |
| Social network | 1 | 9 |
| Other | 2 | 8 |

| 7. Aid tools and linguistic resources | | |
|---|---|---|
| | Yes | No |
| Monolingual Dictionaries | 3 | 7 |
| Multilingual Dictionaries | 4 | 6 |
| Natural language processing tools | 2 | 8 |
| Search engines | 2 | 8 |
| e-books | 1 | 9 |
| Links to other websites | 2 | 8 |

| 8. Website Ergonomics | | | | | |
|---|---|---|---|---|---|
| | 1 Very poor | 2 Poor | 3 Medium | 3 Good | 4 Very good |
| Color harmony | 0 | 1 | 8 | 1 | 0 |
| Font and legibility | 0 | 1 | 8 | 1 | 0 |
| General Structure and Organization | 0 | 1 | 8 | 1 | 0 |
| Ease of Navigation | 0 | 1 | 8 | 1 | 0 |
| Loading speed | 0 | 1 | 8 | 1 | 0 |

| 9. Content Quality | | | | | |
|---|---|---|---|---|---|
| | 1 Very poor | 2 Poor | 3 Medium | 3 Good | 4 Very Good |
| Usefulness | 0 | 0 | 9 | 1 | 0 |
| Accuracy | 0 | 0 | 9 | 1 | 0 |
| Adequacy | 0 | 0 | 9 | 1 | 0 |
| Reliability | 1 | 8 | 0 | 1 | 0 |